\documentclass[aps,prb,twocolumn,nobibnotes]{revtex4-1}  
\usepackage{dcolumn}
\usepackage{graphicx}
\usepackage{color}
\usepackage{amssymb}
\usepackage{amsmath}
\usepackage{braket}
\usepackage{bm}
\usepackage[colorlinks=true,linktocpage=true,breaklinks=true]{hyperref}

\DeclareUnicodeCharacter{2212}{-}

\usepackage{siunitx}
\usepackage{textcomp}

\begin{document}

\title{Adiabatic Spin and Orbital Pumping in Metallic Heterostructures}

\author{Armando Pezo}
\affiliation{Aix-Marseille Univ, CNRS, CINaM, Marseille, France}
\author{Dongwook Go}
\affiliation{Peter Grünberg Institut and Institute for Advanced Simulation, Forschungszentrum Jülich and JARA, 52425 Jülich, Germany}
\affiliation{Institute of Physics, Johannes Gutenberg University Mainz, 55099 Mainz, Germany}
\author{Yuriy Mokrousov}
\affiliation{Peter Grünberg Institut and Institute for Advanced Simulation, Forschungszentrum Jülich and JARA, 52425 Jülich, Germany}
\affiliation{Institute of Physics, Johannes Gutenberg University Mainz, 55099 Mainz, Germany}
\author{Henri Jaffrès} 
\affiliation{Laboratoire Albert Fert, CNRS, Thales, Université Paris-Saclay, 91767, Palaiseau, France}
\author{Aur\'{e}lien Manchon}
\affiliation{Aix-Marseille Univ, CNRS, CINaM, Marseille, France}
\email{aurelien.manchon@univ-amu.fr}

\begin{abstract}
In this study, we investigate the spin and orbital densities induced by magnetization dynamics in a planar bilayer heterostructure. To do this, we employed a theory of adiabatic pumping using the Keldysh formalism and Wigner expansion. We first conduct simulations on a model system to determine the parameters that control the spin and orbital pumping into an adjacent non-magnetic metal. We conclude that, in principle, the orbital pumping can be as significant as spin pumping when the spin-orbit coupling is present in the ferromagnet. We extend the study to realistic heterostructures involving heavy metals (W, Pt, Au) and light metals (Ti, Cu) by using first-principles calculations. We demonstrate that orbital pumping is favored in metals with $d$ states close to the Fermi level, such as Ti, Pt, and W, but is quenched in materials lacking such states, such as Cu and Au. Orbital injection is also favored in materials with strong spin-orbit coupling, leading to large orbital pumping in Ni/(Pt, W) bilayers.
\end{abstract}

\maketitle

\section{Introduction}

Spin currents are routinely used to transmit spin information in nonlocal magnetic spin valves \cite{Jedema2001} and convey spin torque in non-volatile memories and nano-oscillators \cite{Brataas2012}. Because of their potential for applications, substantial efforts have been devoted over the past two decades to generating large, charge-neutral, pure spin currents. This can be achieved using ferromagnetic metals (via direct injection or lateral spin diffusion \cite{Jedema2001}), or materials and heterostructures with strong spin-orbit coupling (via, e.g., the spin Hall effect \cite{Sinova2015} or the Rashba-Edelstein effect \cite{Manchon2015}). Alternatively, one can also harvest the periodic precession of a resonantly excited magnetization, leading to the adiabatic spin pumping \cite{Brataas2002,Tserkovnyak2002,Tserkovnyak2002b,Mizukami2002,Azevedo2005,Saitoh2006}. Spin pumping, which is now a foundational concept of modern spintronics, has been investigated in a wide variety of magnetic interfaces, involving transition metal ferromagnets \cite{Saitoh2006}, ferrimagnetic insulators \cite{Sandweg2011} and antiferromagnets \cite{Vaidya2020}.

Over the past few years, currents of orbital angular momentum, called orbital currents, have emerged as a potential alternative to spin currents \cite{Bernevig2005b,Go2017}. Orbital currents present several remarkable features compared to spin currents. Because the orbital degree of freedom is governed by the crystal field rather than the magnetic exchange or spin-orbit coupling, orbital currents can be generated very efficiently via the orbital Hall effect \cite{Tanaka2008,Jo2018,Bhowal2020,Cysne2021,Pezo2022,Pezo2023} and the orbital Rashba-Edelstein effect \cite{Go2017,Yoda2018,Go2020,Go2020b}, even in the absence of spin-orbit coupling. In addition, once injected in a ferromagnet, the relaxation length of orbital currents tends to exceed that of spin currents \cite{Lee2021b,Hayashi2023}, leading to enhanced orbital torques in metallic multilayers \cite{Ding2020,Lee2021b,Ding2022,Sala2022,Hayashi2023,Fukunaga2023,Lyalin2023,Choi2023}. 

Several theoretical investigations have been proposed to clarify the transport properties (diffusion, relaxation) of pure orbital currents \cite{Go2020,Go2020b,Han2022,Ning2023,Ovalle2023b,Go2023,Sohn2024}, as well as the interplay between orbital and spin currents mediated by spin-orbit coupling. As a matter of fact, in the presence of spin-orbit coupling, a pure spin current is accompanied by a pure orbital current and vice-versa \cite{Ning2023}. Spin pumping is therefore naturally accompanied by orbital pumping when spin-orbit coupling is present. Orbital pumping has been recently investigated theoretically \cite{Han2023,Go2023} and reported experimentally in Ti/Fe and Ti/Ni bilayers \cite{Hayashi2024}, as well as in THz emission measurements \cite{Seifert2023}.

In this article, we investigate dc orbital and spin pumping in realistic metallic bilayers using first-principles calculations. We assess the spin and orbital pumping efficiencies in selected interfaces of interest to experiments. Our work is organized as follows: In section \ref{s:2}, we briefly expose the methodology based on the Keldysh formalism and recently discussed in Ref. \onlinecite{Manchon2024}. In section \ref{s:3}, we present our results on a model bilayer system, followed by realistic $\it{ab}$ $\it{initio}$ simulations on selected bilayers in Section \ref{s:4}. Finally, in Section \ref{s:5}, we discuss the implications of our findings and provide some perspectives.

\section{Methodology\label{s:2}}
 \subsection{Summary of the adiabatic pumping formalism}
Different formalisms have been used to describe spin pumping in magnetic multilayers. The original theory was derived using the scattering matrix approach and soon extended to nonequilibrium Green's function techniques that better account for the effect of spin-orbit coupling \cite{Mahfouzi2010,Mahfouzi2012,Chen2015c,Ly2022}. The two recent theoretical studies on orbital pumping, Refs. \onlinecite{Han2023} and \onlinecite{Go2023}, also consider nonequilibrium Green's functions but applied to very different situations. In Ref. \onlinecite{Han2023}, the authors investigate orbital pumping induced by either magnetic dynamics or lattice dynamics. The source of the orbital current is a metallic layer, described by an effective tight-binding model with $p$-orbitals embedded between two leads. In contrast, Ref. \onlinecite{Go2023} focuses on the orbital currents pumped by a precessing magnetization in the bulk of realistic 3$d$ transition metal ferromagnets. Hence, the former study is more transparent, uncovering the microscopic mechanisms responsible for the various ac and dc orbital currents, while the latter is more realistic but overlooks the role of interfaces. In the present work, we exploit the formalism introduced in Ref. \onlinecite{Manchon2024}, and that is well adapted to treat realistic metallic multilayers. This way, we can evaluate the actual injection of spin and orbital currents from the ferromagnet into the nonmagnetic metal. We summarize this formalism below.

 \subsection{Keldysh approach to adiabatic pumping}
Following Ref. \onlinecite{Manchon2024}, we consider a multi-orbital system defined by the Hamiltonian ${\cal H}(t)$ that is time-dependent. In the case of a magnetic heterostructure with a precessing magnetization, the Hamiltonian depends explicitly on the magnetization ${\bf m}(t)$ and one can define a torque operator ${\bf T}$ such that $\partial_t{\cal H}=-{\bf T}\cdot\partial_t{\bf m}$, or, equivalently, ${\bf T}=-\partial_{\bf m}{\cal H}$. In this case, an observable ${\cal O}_i$ can be expressed at the first order in time-derivative of the magnetization as ${\cal O}_i=\sum_j\delta{\cal O}_{ij}\partial_tm_j$, where $\delta{\cal O}_{ij}$ is an element of the linear response tensor, given by
\begin{eqnarray}
 \delta{\cal O}_{ij}^{sym}=-\hbar \int \frac{d\varepsilon}{4\pi}{\rm ReTr}[{\cal O}_iG^{R-A}T_jG^{R-A}]\partial_{\varepsilon} f(\varepsilon), \label{eq:sym}\\
\delta{\cal O}_{ij}^{asym}=\hbar \int \frac{d\varepsilon}{2\pi}{\rm ReTr}[{\cal O}_iG^{R+A}T_j\partial_{\varepsilon}G^{R-A}]f(\varepsilon).
    \label{eq:asym}
\end{eqnarray}
Here, $G^{R\pm A}=G^{R}\pm G^{A}$. The two contributions are the symmetric part, $\delta{\cal O}_{ij}^{sym}=\delta{\cal O}_{ji}^{sym}$, and antisymmetric part, $\delta{\cal O}_{ij}^{asym}=-\delta{\cal O}_{ji}^{asym}$, of the response tensor. The former, Eq. \eqref{eq:sym}, is a Fermi surface contribution that is even in magnetization whereas the latter, Eq. \eqref{eq:asym}, is a Fermi sea term that is odd in magnetization. As a result, Eq. \eqref{eq:sym} represents the ac pumping whereas Eq. \eqref{eq:asym} produces both ac and dc contributions \cite{Manchon2024}. This theory is applicable as long as the frequency of the magnetization precession is small compared to the exchange energy. Extending this theory to pumping by lattice vibrations is readily possible, provided that one defines a coupling matrix that explicitly accounts for the time-dependence of the atomic positions, as in Ref. \onlinecite{Han2023}. Notice that a difference with Ref. \onlinecite{Go2023} is that we compute the spin and orbital densities, $\langle{\bf s}\rangle$ and $\langle{\bf l}\rangle$, whereas Ref. \onlinecite{Go2023} computes the rates $\partial_t\langle{\bf s}\rangle$ and $\partial_t\langle{\bf l}\rangle$. \par 

Since we are interested in the generation of (spin and orbital) dc currents, we focus on Eq. \eqref{eq:asym} in the rest of the study. This equation can be cast in terms of the Bloch functions by expanding the Green's functions based on the eigenvalues and eigenstates of $\mathcal{H}$,
\begin{equation}
     G_0^{\rm R(A)}(\varepsilon)=\sum_n\frac{\ket{u_{n\mathbf{k}}}\bra{u_{n\mathbf{k}}}}{\varepsilon-\varepsilon_{n\mathbf{k}}\pm i\Gamma},\label{bloch}
\end{equation}
where $\varepsilon_{n\mathbf{k}}$ is the eigenvalue associated with the eigenstate  $\ket{u_{n\mathbf{k}}}$, and $\Gamma$ is a homogeneous broadening arising from impury scattering. Notice that in the present work, self-consistent vertex corrections are disregarded as they are numerically prohibitive for realistic Hamiltonians. By using Eq. \eqref{bloch}, we avoid a time-consuming energy integration that is replaced by the following expression

\begin{eqnarray}
    \delta\mathcal{O}_{ij}&=&-\frac{\hbar}{\pi}\sum_{n,m\neq n}\Im \{ \braket{u_{m\mathbf{k}}|\mathcal{O}_i|u_{n\mathbf{k}}}\braket{u_{n\mathbf{k}}|T_j|u_{m\mathbf{k}}} \}\\
    \label{bloch_2}
    &&\times \int d\varepsilon \left\{\left(\frac{1}{\varepsilon-\varepsilon_m-i\Gamma}+\frac{1}{\varepsilon-\varepsilon_m+i\Gamma}\right)\right.\nonumber\\
   && \times \left.\left(\frac{1}{(\varepsilon-\varepsilon_n-i\Gamma)^2}- \frac{1}{(\varepsilon-\varepsilon_n+i\Gamma)^2}\right)\right\} f(\varepsilon).\nonumber
\end{eqnarray}

By integration by parts, we get
\begin{equation}
    \delta\mathcal{O}_{ij}(\varepsilon)=-\frac{\hbar}{\pi}\sum_{n,m\neq n}\Im \{ \braket{u_{m\mathbf{k}}|\mathcal{O}_i|u_{n\mathbf{k}}}\braket{u_{n\mathbf{k}}|T_j|u_{m\mathbf{k}}} \}G(\varepsilon),
    \label{bloch_final}
\end{equation}
where

\begin{equation}
G(\varepsilon)= \frac{\Gamma}{2(\varepsilon_n-\varepsilon_m)^2}\bigg(\frac{\varepsilon_n-\varepsilon_m}{4\Gamma^2+(\varepsilon_n-\varepsilon_m)^2} \times
\end{equation}
\begin{equation*}
\Bigl\{ \frac{2\Gamma^2+(\varepsilon_n-\varepsilon_m)(\varepsilon-2\varepsilon_m-\varepsilon_n)}{(\varepsilon-\varepsilon_m)^2+\Gamma^2}
\end{equation*}
\begin{equation*}
+\frac{2\Gamma^2-(\varepsilon_n-\varepsilon_m)(\varepsilon-2\varepsilon_n+\varepsilon_m)}{(\varepsilon-\varepsilon_n)^2+\Gamma^2}\Bigl\}
\end{equation*}
\begin{equation*}
+\frac{1}{\Gamma}(\arctan{\frac{\varepsilon-\varepsilon_n}{\Gamma}}-\arctan{\frac{\varepsilon-\varepsilon_m}{\Gamma}})
\bigg).
\end{equation*}

 \subsection{Procedure to extract dc pumping effects} 
We focus on the dc response and therefore look for terms that do not vanish upon time-averaging over one period of magnetization precession. Let us consider the most experimentally relevant situation \cite{Saitoh2006}, where the magnetization precesses around the direction ${\bf y}$, i.e., ${\bf m}=\cos\theta{\bf y}-\sin\theta(\sin\omega t{\bf x}+\cos\omega t{\bf z})$, $\theta$ being the cone angle and $\omega$ the precession frequency. Writing $\delta{\cal O}_{ij}^{asym}=\sum_l\chi_{jl}^im_l$, the dc response reads
 \begin{eqnarray}
 {\cal O}_i=\sum_{jl}\chi_{jl}^i\langle m_l\partial_tm_j\rangle=\frac{\omega}{2}(\chi^i_{xz}-\chi^i_{zx})\sin^2\theta,
 \end{eqnarray}
 where $\langle...\rangle$ indicates time averaging over the precession period. Therefore, the procedure boils down to computing the response functions
 \begin{eqnarray}\label{eq:resp1}
\chi^i_{xz}=\frac{1}{m_x}\delta{\cal O}^{asym}_{iz},\;\chi^i_{zx}=\frac{1}{m_z}\delta{\cal O}^{asym}_{ix}.\label{eq:resp4}
\end{eqnarray}
These values are computed at the four points of the precession circle, ${\bf m}=\cos\theta{\bf y}\pm\sin\theta{\bf x}$ and ${\bf m}=\cos\theta{\bf y}\pm\sin\theta{\bf z}$ for each calculation.\\

\section{Orbital pumping in a model bilayer\label{s:3}}
\subsection{Definition of the system}
Let us first consider a model system composed of two layers, one ferromagnetic (FM) and one nonmagnetic metal (NM), with a cubic unit cell and involving $p$ and $d$ orbitals. The Hamiltonian of the bilayer reads

\begin{equation}
\hat{\cal H} = \begin{pmatrix}
\hat{\cal H}_{\rm FM} & \hat{\cal H}_{\rm C} \\
\hat{\cal H}^\dagger_{\rm C} & \hat{\cal H}_{\rm NM}\\
\end{pmatrix}\label{eq:big_hamiltonian},
\end{equation}
where $\hat{\cal H}_{\rm FM/NM}$ are the Hamiltonian matrices corresponding to the FM and NM sides and $\hat{\cal H}_{\rm C}$ is the coupling between the two layers. To construct the slabs, we start by considering a standard Slater-Koster parameterization, such that the unit cell has always a lattice parameter of $a=1$ and each bulk Hamiltonian can be written as
 \begin{equation}
     \hat{\cal H}^{\rm FM/NM}_{bulk}=\sum_{\alpha \beta,\sigma}t^{\rm FM/NM}_{\alpha \beta \sigma}\hat{c}^{\dagger}_{\alpha \sigma}\hat{c}_{\beta \sigma}+\lambda^{\rm FM/NM}_{\rm soc}\hat{\mathbf{L}}\cdot\hat{\bm\sigma},
 \end{equation}
where the hopping matrix elements $t^{\rm FM/NM}_{\alpha \beta \sigma}$ are set to the same value within each system. Here, $\sigma$ is the spin index, and $\alpha,\beta$ are the orbital indices. $\lambda^{\rm FM/NM}_{\rm soc}$ represents the strength of the spin-orbit coupling of layer FM or NM, with $\hat{\bm\sigma}$ the spin momentum operator and $\hat{\mathbf{L}}$ the usual orbital angular momentum matrices for $p$ and $d$ states. In addition, the FM layer possesses an exchange interaction given by
\begin{equation}
    \hat{\cal H}^{\rm exc}_{bulk}=\Delta\hat{I}\otimes \hat{\bm\sigma}\cdot\mathbf{m}.
\end{equation}
where $\hat{I}$ is the identity in orbital space.
\begin{figure}[!ht]
    \centering    \includegraphics[width=\linewidth]{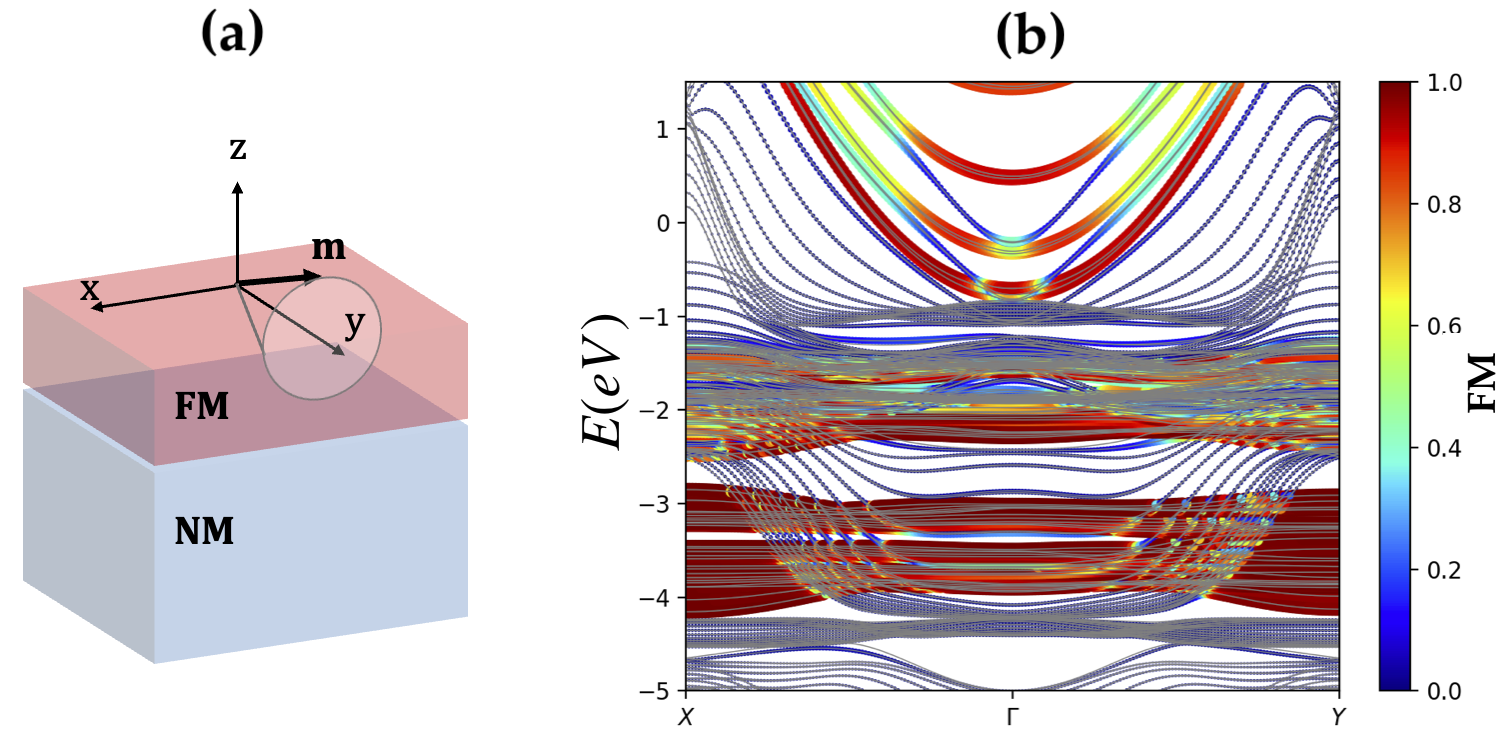}
    \caption{(Color online) (a) Schematics of the bilayer with a magnetization precessing around the $y$ axis. (b) Band structure of the bilayer projected on the ferromagnetic layer. The color scale indicates the layer character of the band: blue corresponds to NM and red corresponds to FM.}
    \label{fig:pumping_bands_model}
\end{figure}
The band structure of the heterostructure is shown in Fig. \ref{fig:pumping_bands_model}. The color code shows the projection on the FM layer. From the band structure, we see that the FM states lie mostly within the energy window $-4$ eV to $-3$ eV such that we set the Fermi level in the middle of this range. On the other hand, the projection coming from the NM layer seems more dispersive with some states entangled around the chosen energy window.

\subsection{Spin and orbital pumping}

\begin{figure*}[!ht]
    \centering
    \includegraphics[width=\linewidth]{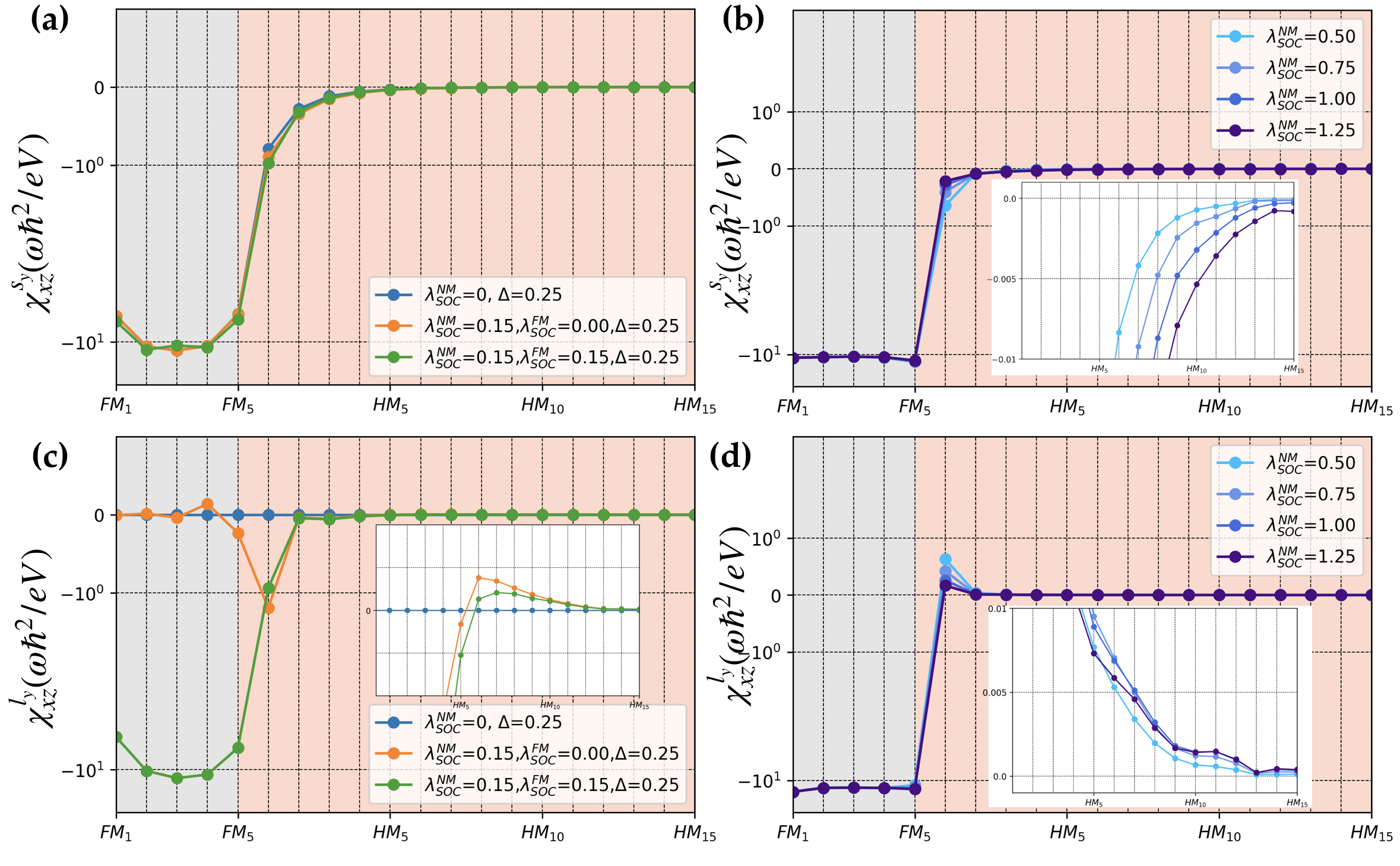}
    \caption{(a,b) Spin and (c,d) orbital density profile across the FM/NM bilayer, (a,c) for $t^{\rm FM}=t^{\rm NM}$ and (b,d) $t^{\rm FM}=t^{\rm NM}/3$, as explained in the text. The grey (pink) region designates the FM (NM) layer. (a) Spin and (c) orbital density profile for $t^{\rm FM}=t^{\rm NM}$. In (b,d), $\lambda_{\rm soc}^{\rm FM}=0.15t$, and the insets show a zoom of the spin and orbital density decay in the NM layer. The exchange parameter is fixed to $\Delta=0.25t$ for all the calculations.}\label{fig:pumping_model_1}
\end{figure*}

For the out-of-equilibrium simulations, we compute Eq. \eqref{bloch_2} sampling the Brillouin zone with a $200 \times 200$-point grid. We first compute the nonequilibrium spin density (${\cal O}_i=s_y$) across the slab as a function of the position, see Fig. \ref{fig:pumping_model_1}(a). In this calculation, the hopping integral is the same for FM and NM layers, $t=0.5$, the onsite energy is taken at $-1.5t$ and the exchange is fixed to $\Delta=0.25t$. Turning on and off the spin-orbit coupling of the FM and NM layers, one immediately sees that the spin density profile is mostly independent of the spin-orbit coupling and quickly vanishes as it penetrates inside the NM layer. In contrast, the orbital density profile (${\cal O}_i=l_y$) reported in Fig. \ref{fig:pumping_model_1}(c) is highly sensitive to the spin-orbit coupling and vanishes in its absence (blue symbols). When the spin-orbit coupling is turned on in both FM and NM layers, the orbital density closely follows the profile of the spin density (green symbols), whereas when the spin-orbit coupling is only present in the NM layer, the spin-to-charge conversion takes place only close to the FM/NM interface (orange symbols). Notice that with our parameters, such a strong spin-orbit coupling leads to a magnitude of the orbital density that is similar to that of the spin density.

Taking different values for the hopping matrix elements of the NM and FM layers, $t^{\rm FM}=t^{\rm NM}/3$, slightly modifies the overall picture, as shown in Figs. \ref{fig:pumping_model_1}(b) and (d). For a finite value of the spin-orbit coupling in the FM layer, $\lambda_{\rm soc}^{FM}=0.15t$, the spin pumping in Fig. \ref{fig:pumping_model_1}(b) is accompanied by orbital pumping in Fig. \ref{fig:pumping_model_1}(c). As shown in the insets of these figures, the decay length of the spin and orbital densities is influenced by the spin-orbit coupling strength of the NM layer. Notice that in the absence of vertex corrections, our calculations do not account for spin relaxation, and the decay of the spin and orbital densities is simply due to the spreading of the electronic wave function out of equilibrium. Therefore, the slight increase in the decay length observed in the insets upon increasing the spin-orbit coupling is associated with the enhancement of the spin-orbit mixing in the NM layer. One interesting aspect we wish to point out is the slight peak of orbital density at the FM/NM interface, shown in Fig. \ref{fig:pumping_model_1}(c). This local enhancement corroborates the presence of the interfacial Rashba effect, as previously studied in Ref. \onlinecite{Chen2015c}, which enhances the spin-to-orbit conversion locally.

\section{Realistic simulations\label{s:4}}
\subsection{Methodology}
Let us now investigate spin and orbital pumping in realistic bilayers combining transition FM (Fe, Co, Ni) and selected NM metals with both weak (Ti, Cu) and strong spin-orbit coupling (Au, Pt, W). 5$d$ Pt and W display a large spin Hall effect leading to strong spin-charge interconversion; one can, therefore, reasonably expect that spin-to-orbital conversion should be large in these materials. 3$d$ Ti is one of the most efficient orbital-charge converters and has already been studied in orbital pumping experiments \cite{Hayashi2024}. 3$d$ Cu and 5$d$ Au possess weak and strong spin-orbit coupling, respectively, with a filled $d$-shell which {\em a priori} prevents orbital-charge and spin-charge interconversion.

We determined the band structures and spin textures of the bilayers by employing fully relativistic density functional theory (DFT) with \textsc{VASP} package \cite{vasp1,vasp2} using a plane-wave basis with the Perdew-Burke-Ernzerhof (PBE) \cite{gga,pbe} exchange-correlation functional. We force the slabs to maintain the lattice constants of the NM layer, leading to a mismatch concerning the actual values of the lattice parameters on the FM. This approximation has been proven to work fairly well \cite{Manchon2020} and its validity and limitations are commented on in the following. We used \SI{400}{\electronvolt} for the plane-wave expansion cutoff and ionic potentials are described using the projector augmented-wave (PAW) method \cite{paw} performing the geometry optimization with a force criterion of \SI{1e-3}{\electronvolt\per\angstrom}. We describe the spin-orbit coupling within a fully relativistic pseudo-potential formulation and use the generalized gradient approximation (GGA) for the exchange-correlation functional, the calculations are converged for a 400 Ry plane-wave cutoff for the real-space grid with a  $(13 \times 13 \times 13)$  $\vec{k}$-points sampling of the Brillouin zone. The spin and orbital pumping responses are calculated based on Eq. \eqref{bloch_2} using a home-made procedure after extracting the Hamiltonian matrices from {\it{ab initio}} using SISL as a post-processing tool.

\subsection{Spin and orbital density profile}

\begin{figure*}[!ht]
    \centering
    \includegraphics[width=\linewidth]{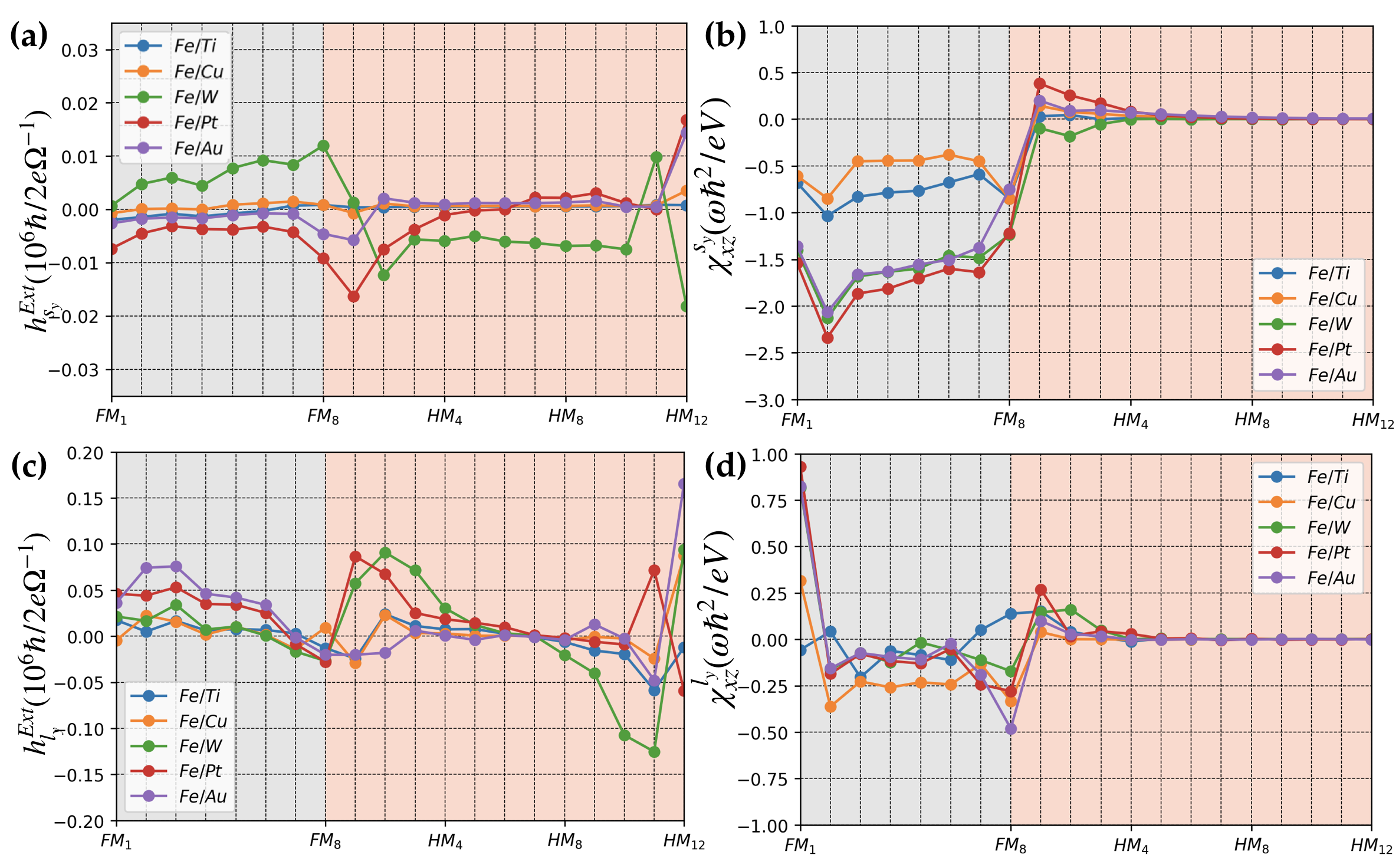}
    \caption{(Color Online) (a,c) Current-driven and (b,d) spin-pumping-driven (a,b) spin and (c,d) orbital density profile in Fe/NM bilayers computed from first principles as described in the main text. In the current-driven case, the orbital densities are about one order of magnitude larger than the spin densities, whereas for the spin-pumping case, it is the opposite.}
    \label{fig:sot_fe}
\end{figure*}

We first compute the nonequilibrium spin and orbital densities in Fe/NM bilayers, displayed in Figs. \ref{fig:sot_fe}. Keeping in mind that orbital pumping is a consequence of spin-orbit coupling, let us first probe the spin and orbital densities induced by a charge current. Fig. \ref{fig:sot_fe}(a,c) shows the (a) spin and (c) orbital densities, $h^{Ext}_{s_y}$ and $h^{Ext}_{l_y}$, obtained when an electric field is applied along the $\hat{x}$ direction and the magnetization is set along $\hat{z}$. To do so, we use the conventional Kubo-Bastin formula \cite{Bonbien2020},
\begin{eqnarray}
  h_{s_y}^{Ext}=-\frac{e\hbar}{8\pi}\int \partial_\varepsilon f(\varepsilon)d\varepsilon \operatorname{Re}\operatorname{Tr}
   \{v_xG^{\rm R-A} \hat{s}_yG^{\rm R-A}\},&&\\
  h_{l_y}^{Ext}=-\frac{e\hbar}{8\pi}\int \partial_\varepsilon f(\varepsilon)d\varepsilon \operatorname{Re}\operatorname{Tr}
   \{v_xG^{\rm R-A} \hat{l}_yG^{\rm R-A}\}, &&  
\end{eqnarray}
with $v_x=\partial_{k_x} \mathcal{H}$ being the velocity operator, and $s_y$, $l_y$ are the spin and orbital angular momentum operators projected on each layer leading to the spin/orbital accumulation on each layer. Analogously to the previous section, the Brillouin zone integration was carried out on a $200 \times 200$ mesh grid.\\

We find that the current-driven spin density [Fig. \ref{fig:sot_fe}(a)] is about one order of magnitude smaller than the orbital density [Fig. \ref{fig:sot_fe}(c)], which is reasonable considering that the current-driven spin density arises from spin-orbit coupling (spin Hall and spin Edelstein effects) whereas the current-driven orbital density does not. As expected, the spin density is much larger in bilayers involving 5$d$ metals (W, Pt, and Au) than in bilayers with 3$d$ metals (Ti and Cu). Notice the opposite sign of the spin density for Pt and W, which is corroborated by experiments and associated with the opposite sign of the spin Hall effect in these two materials \cite{Manchon2019}. In contrast, the orbital density is about the same order of magnitude across the nonmagnetic metals, with Fe/(W, Pt) giving the largest value and Fe/(Ti, Cu) the smallest. These results are qualitatively consistent with previous evaluations of the spin and orbital Hall effect in bulk transition metals \cite{Salemi2022}. We emphasize that the spin and orbital density profiles reported in Fig. \ref{fig:sot_fe}(a,c) arise from the cooperation between spin and orbital Hall effects in the materials' volume, and spin and orbital Edelstein effect at the three (two outer and one inner) interfaces. In the case of Au for instance, the 5$d$ shell is filled, resulting in a very small spin Hall effect, but a Rashba state appears at its surface, giving rise to a large interfacial spin density.
\begin{figure*}[t!]
    \centering
    \includegraphics[width=\linewidth]{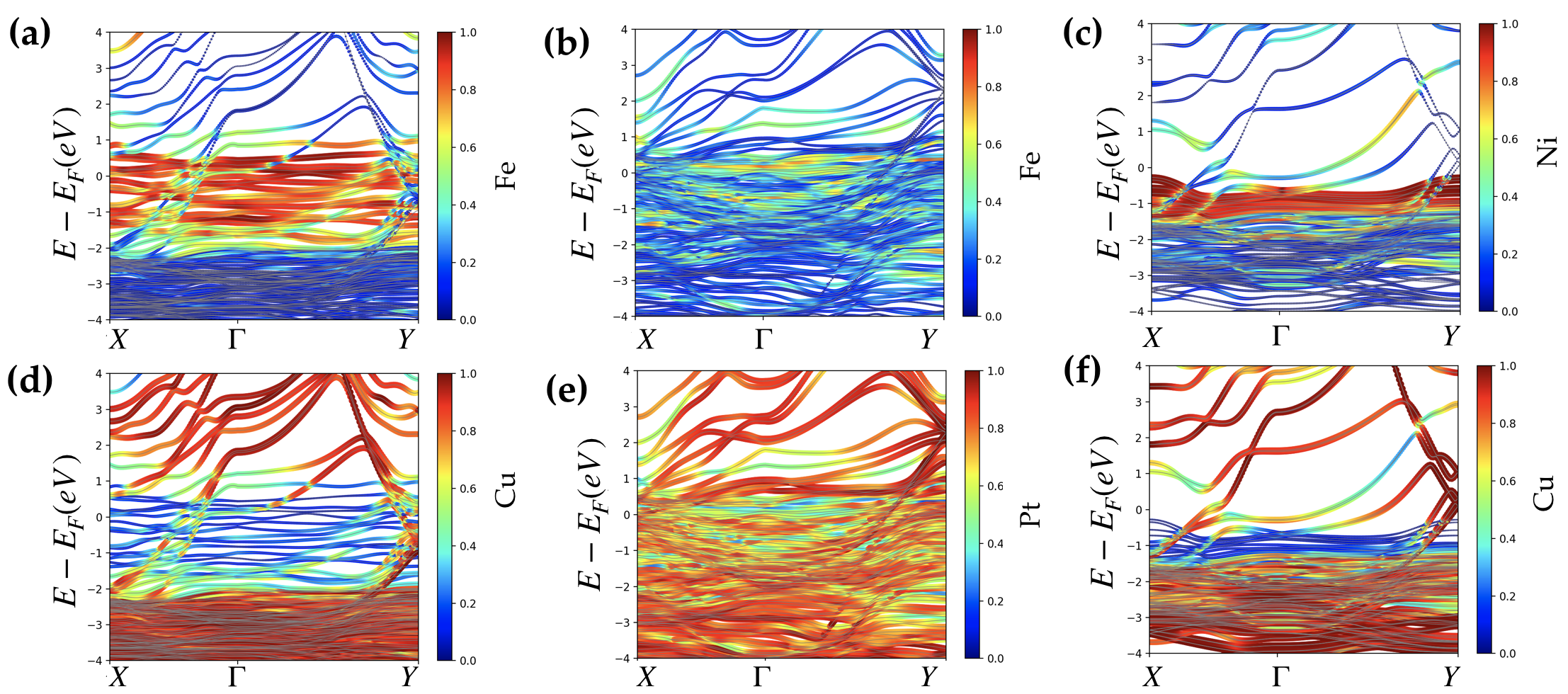}
    \caption{(Color online) Band structure for the (a,d) Fe/Cu, (b,e) Fe/Pt and (c,f) Ni/Cu bilayers projected on the FM layer (a,b,c) and the NM layer (d,e,f).}
    \label{fig:bands_comb}
\end{figure*}

We now consider the nonequilibrium spin and orbital densities induced by spin pumping, displayed in Fig. \ref{fig:sot_fe}(b,d). The picture is radically different. In Fig. \ref{fig:sot_fe}(b), we find that the spin density in the FM layer (grey region) is much larger for Fe/(W, Pt, Au) than for Fe/(Ti, Cu). More interestingly, we find that on the NM layer (pink region), spin pumping is more efficient into 5$d$ metals than into 3$d$ systems, which corroborates experimental observations. In an attempt to explain these results, the band structure of Fe/Cu and Fe/Pt projected on the FM (a,b) and NM layers (d,e) are reported in Figs. \ref{fig:bands_comb}. In Fe/Cu, the Fe states are located around the Fermi level, whereas Cu states are located far from it, reducing spin injection from Fe to Cu. In contrast, in Fe/Pt, the Fe states are entangled with the Pt states around the Fermi level, favoring spin injection from Fe into Pt. 

The orbital density induced by magnetization precession is shown in Fig. \ref{fig:sot_fe}(d). The orbital density in the FM layer is very sensitive to the interface akin to the spin density and displays an enhancement at the left surface, likely associated with the interfacial Rashba effect that favors spin-orbital conversion. In the NM layer, the orbital density profile exhibits an instructive feature: its decay length is much longer in Pt, W, and Ti than in Au and Cu. In other words, the injection of the orbital moment in a metal is favored by the presence of $d$-orbital at the Fermi level. In metals whose Fermi electrons possess mainly an $s$-character, such as Au and Cu, the injection of the orbital moment is quenched.

\begin{figure*}[t]
    \centering    \includegraphics[width=\linewidth]{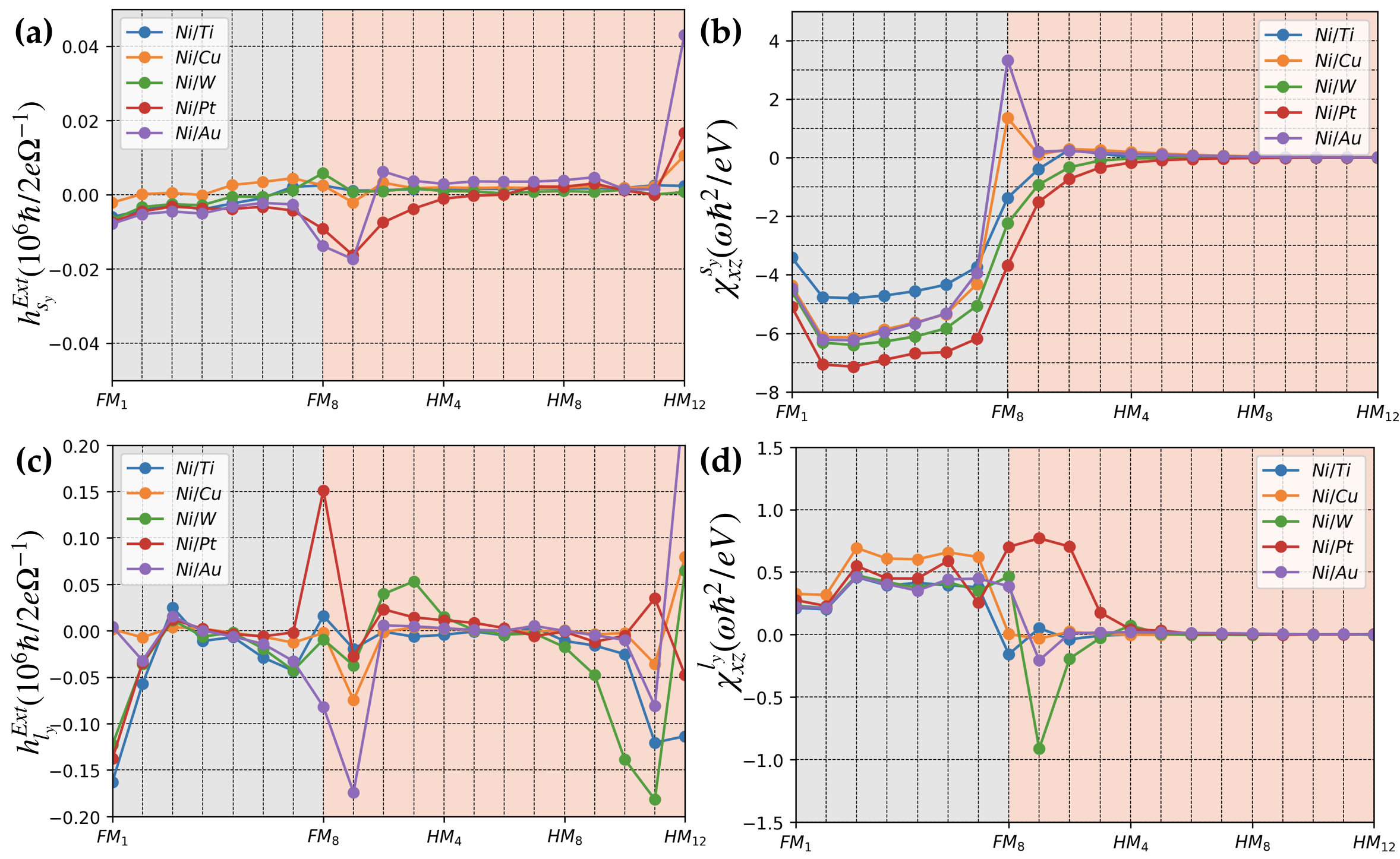}
    \caption{(Color Online) (a,c) Current-driven and (b,d) spin-pumping-driven (a,b) spin and (c,d) orbital density profile in Ni/NM bilayers computed from first principles as described in the main text. Similar features are observed, with very large orbital pumping in Ni/Pt and Ni/W bilayers.}
    \label{fig:pumping_Ni}
\end{figure*}

Let us now consider the nonequilibrium spin and orbital densities in Ni-based bilayers, reported in Fig. \ref{fig:pumping_Ni}. The overall trends of current-driven spin and orbital densities, displayed in Figs. \ref{fig:pumping_Ni}(a) and (b), respectively, are similar to those found in Fe-based bilayers. The current-driven orbital density profile is generally much larger than the current-driven spin density, and the spin density is generally larger in bilayers involving 5$d$ metals than 3$d$ metals. Remarkably, the current-driven orbital density shows large interfacial enhancement in Ni/Pt and Ni/Au, as well as at the outer surfaces of the bilayers, associated with the interfacial Rashba effect. Spin and orbital pumping are reported in Figs. \ref{fig:pumping_Ni}(b,d), respectively. The nonequilibrium spin density profile is smoother and much larger in Ni-based bilayers than in Fe-based bilayers, and, again, the largest spin injections are obtained for Ni/(Pt,W), with local enhancement at the Ni/Au interface. In Fig. \ref{fig:pumping_Ni} (d), we observe that the orbital injection is again much for efficient into W and Pt (with opposite sign) due to the combination of the high spin injection and a strong spin-orbit coupling. Similarly to the case of Fe/NM bilayers, orbital injection is generally more efficient in metals with $d$ states at the Fermi level.

\subsection{Rationalizing spin and orbital injection}

\begin{figure}[!ht]
    \centering
    \includegraphics[width=\linewidth]{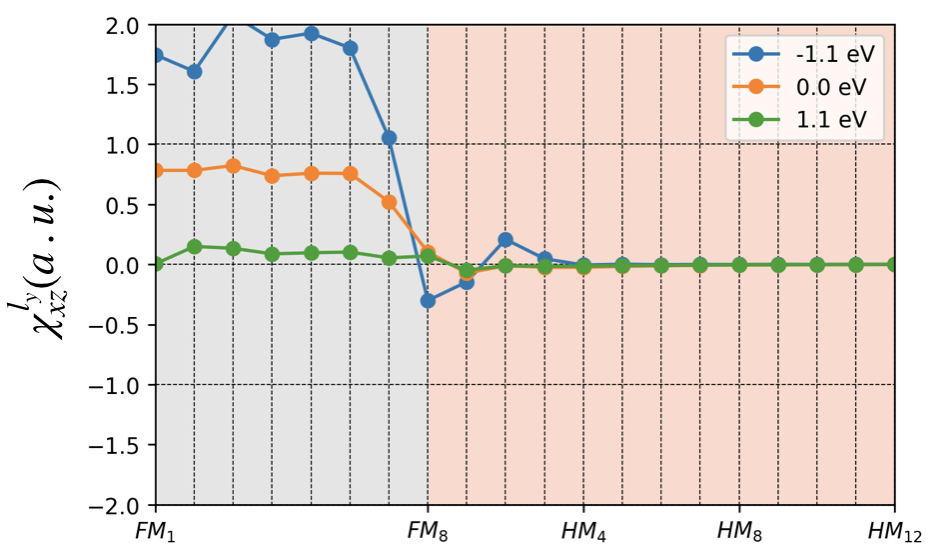}
    \caption{Orbital pumping in Ni/Cu for three different chemical potential values. At Fermi level (orange) and above (green), no $d$ states are available in Cu, and orbital injection is quenched. As the chemical potential is set below the Fermi level (blue), $d$ states become available, allowing for the injection of the orbital moment into Cu.}
    \label{fig:comparison_cu_energies}
\end{figure}

Our simulations of spin and orbital pumping reveal differences between these two degrees of freedom. First, it appears that Ni/NM bilayers are more efficient spin and orbital sources than Fe/NM. In addition, the penetration of orbital moment into the NM layer is favored when $d$ states are present close to the Fermi level (Ti,W,Pt) and quenched when no such states are present (Cu,Au). This feature is only observed for orbital pumping, not for spin pumping. To illustrate the importance of the presence of these $d$ states close to the Fermi level, we computed in Fig. \ref{fig:comparison_cu_energies} the orbital density profile in Ni/Cu for two additional transport energies, i.e., above (1.1 eV) and below (-1.1 eV) the Fermi level (taken at 0.0 eV). Clearly, above the Fermi level, no Cu $d$ states are present, and orbital pumping is inefficient, whereas below Fermi energy, not only the orbital density inside Ni is rather larger, but orbital injection into Cu is rather efficient. This calculation mimics the behavior observed in oxidized Cu, where $d$ orbitals are unfilled by the presence of oxygen elements \cite{Go2020}. 

\section{Discussion and Conclusion\label{s:5}}

\begin{figure}[!ht]
    \centering
    \includegraphics[width=\linewidth]{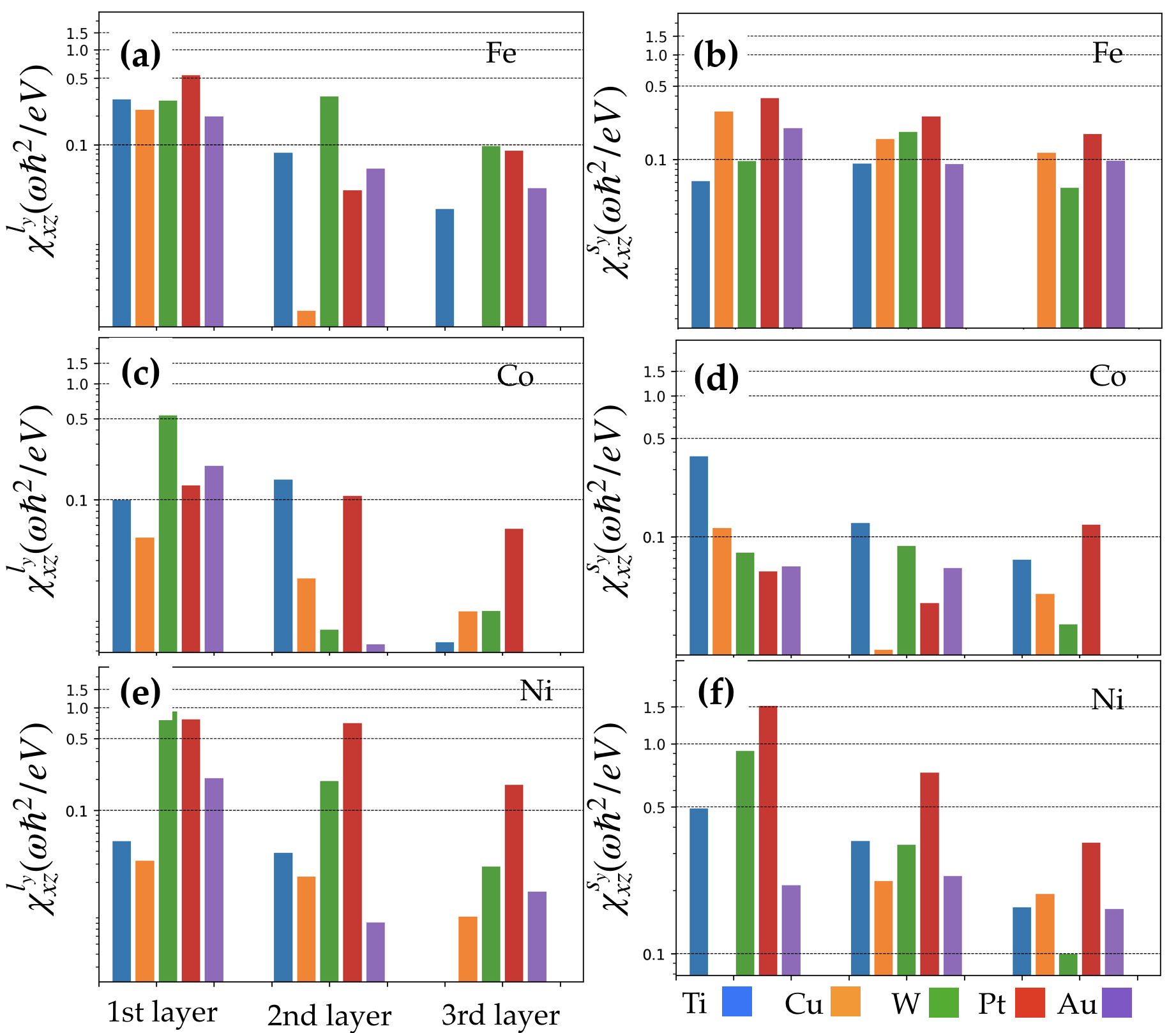}
    \caption{Summary of the (a,c,e) orbital and (b,d,f) spin densities induced by spin pumping in the first three monolayers of the NM layer, for NM=Ti, Cu, W, Pt, and Au. The FM layer is (a,b) Fe, (c,d) Co, and (e,f) Ni. Whereas Fe is a good spin injector for all NM layers considered, the best orbital injectors are Ni/Pt and Fe/W.}
    \label{fig:summary}
\end{figure}

To provide a complete description of spin and orbital pumping in bilayers, we summarize the densities obtained in the first three monolayers of NM in a wide series of FM/NM bilayers, with FM=Fe, Co, Ni and NM=Ti, Cu, W, Pt, and Au. The results are reported in Fig. \ref{fig:summary}. Our simulations reveal that the largest values for the spin pumping correspond to the case of Ni, followed by Fe and Co, which is in qualitative agreement with the recent calculations \cite{Go2023}. In Ref. \onlinecite{Go2023}, it was found that the generation of orbital moment by magnetic pumping increases from about 5 in Fe to 15 \% in Ni compared to spin moment. We emphasize, though, that Ref. \onlinecite{Go2023} focuses on the orbital pumping taking place in the bulk of Fe, Co, and Ni, whereas our approach enables us to consider FM/NM bilayers, where the role of the interface dominates the pumping process. Hence, orbital pumping takes place not only inside the FM layer but also at the interface, where enhanced spin-orbit coupling can lead to enhanced orbital pumping.  This is particularly clear in the case of Ni/Pt and Ni/W, as displayed in Fig. \ref{fig:pumping_Ni}. In general, we find that orbital pumping is more efficient in metals with strong spin-orbit coupling (e.g., Pt, W), as larger spin-orbit interconversion occurs.

Another important feature revealed by our calculations is the role of the orbital nature of the Fermi electrons. We find that for orbital injection to be optimal, the Bloch states present at the Fermi level must carry an orbital degree of freedom, typically $p$ or $d$ states. As a result, orbital injection in Cu or Au is severely quenched because no $d$ states are available. 

Let us comment on the recent experiments performed by Hayashi et al. \cite{Hayashi2024}. In this work, the authors observe that the charge current induced by spin pumping in Ni/Ti largely exceeds that measured in Fe/Ti. Since Ti is known as an efficient orbital-charge converter, this observation is interpreted as a demonstration of orbital pumping from Ni followed by orbital-charge conversion in Ti. Our calculations displayed in Fig. \ref{fig:summary} show that spin pumping is about one order of magnitude more efficient in Ni/Ti than in Fe/Ti, but that the orbital pumping is rather similar in both bilayers. Therefore, our calculation offers an alternative scenario to the one proposed by Hayashi et al. \cite{Hayashi2024}. Since Ti has a small but finite spin-orbit coupling, one could speculate that it is sufficiently large to induce spin-charge conversion. Because spin pumping is much more efficient in Ni/Ti compared to Fe/Ti, the resulting pumped charge current is detectable in Ni/Ti and vanishingly small in Fe/Ti. Testing the different scenarios would require further development of the present calculation to properly include the role of disorder and the spin-charge conversion mechanisms, which is left for future work.

Before concluding, we need to stress two important aspects that were not accounted for in our study and can massively impact spin and orbital pumping. First, as mentioned above, the disorder is accounted for only through the energy broadening $\Gamma$, but does not properly describe diffusive transport. Extending the present study to the diffusive regime is a necessary step toward predictive modeling. Second, orbital transport is expected to be even more sensitive to interfacial roughness and polycrystallinity than spin transport. These aspects are particularly difficult to model from first principles. Nonetheless, our results are encouraging as they shed light on spin and orbital pumping in metallic bilayers and might serve to guide experiments on orbital transport.

\acknowledgments
 
This work was supported by the ANR ORION project, grant ANR-20-CE30-0022-01 of the French Agence Nationale de la Recherche, by the Excellence Initiative of Aix-Marseille Université-A*Midex, a French ”Investissements d’Avenir” program, by France 2030 government investment plan managed by the French National Research Agency under grant reference PEPR SPIN – [SPINTHEORY] ANR-22-EXSP-0009, and by the EIC Pathfinder OPEN grant 101129641 “OBELIX”. D. Go and Y. Mokrousov gratefully acknowledge the J\"ulich Supercomputing Centre for providing computational resources under project jiff40. This work was funded by the Deutsche Forschungsgemeinschaft (DFG, German Research Foundation)−TRR 173−268565370 (project A11), TRR 288−422213477 (project B06).

\bibliography{refs-resub-2}
\end{document}